\documentclass[aip,reprint]{revtex4-2}
\usepackage{graphicx,dcolumn,CJK,xparse,physics,color,bm,ulem}
\usepackage[colorlinks=true,breaklinks=true,allcolors=blue]{hyperref}
\newcommand{\p}[2]{#1^{\text{\tiny($#2$)}}}
\draft 

\begin{document}
\title{Fidelity and quantum geometry approach to Dirac exceptional points in diamond nitrogen-vacancy centers} 

\author{Chia-Yi Ju}
\affiliation{\mbox{Department of Physics, National Sun Yat-Sen University, Kaohsiung 80424, Taiwan}}
\affiliation{\mbox{Center for Theoretical and Computational Physics, National Sun Yat-Sen University, Kaohsiung 80424, Taiwan}}
\affiliation{\mbox{Physics Division, National Center for Theoretical Sciences, Taipei 106319, Taiwan}}

\author{Gunnar M\"oller}
\affiliation{Physics of Quantum \& Materials Group, School of Engineering, Mathematics and Physics, University of Kent, Canterbury CT2 7NH, United Kingdom}\affiliation{Department of ElectroPhysics and Center for Theoretical and Computational Physics, National Yang Ming Chiao Tung University, Hsinchu 300093, Taiwan}

\author{Yu-Chin Tzeng}\email{yctzeng@nycu.edu.tw}
\affiliation{Physics of Quantum \& Materials Group, School of Engineering, Mathematics and Physics, University of Kent, Canterbury CT2 7NH, United Kingdom}\affiliation{Department of ElectroPhysics and Center for Theoretical and Computational Physics, National Yang Ming Chiao Tung University, Hsinchu 300093, Taiwan}

\begin{abstract}
Dirac exceptional points (EPs) represent a novel class of non-Hermitian singularities that, unlike conventional EPs, reside entirely within the parity-time unbroken phase and exhibit linear energy dispersion. Here, we theoretically investigate the quantum geometry of Dirac EPs realized in nitrogen-vacancy centers in diamond, utilizing fidelity susceptibility as a probe. We demonstrate that despite the absence of a symmetry-breaking phase transition, the Dirac EP induces a pronounced geometric singularity, confirming the validity of the fidelity in characterizing non-Hermitian EPs. Specifically, the real part of the fidelity susceptibility diverges to negative infinity, which serves as a signature of non-Hermitian criticality. Crucially, however, we reveal that this divergence exhibits a distinct anisotropy, diverging along the non-reciprocal coupling direction while remaining finite along the detuning axis. Furthermore, we establish that this anisotropy, characterized by at least one exact dark direction coexisting with divergent directions, is a generic consequence of the Dirac EP structure whenever the parameter derivatives collectively span the off-diagonal operator space at the Dirac EP. This behavior stands in stark contrast to the omnidirectional divergence observed in conventional EPs. Our findings provide a comprehensive picture of the fidelity probe near the Dirac EP, highlighting the critical role of parameter directionality in exploiting Dirac EPs for quantum control and sensing applications.
\end{abstract}


\maketitle 

\section{Introduction}
Non-Hermitian physics has emerged as a fertile ground for discovering exotic topological features and singularities that have no counterparts in Hermitian systems. At the heart of these unique phenomena lie the exceptional points (EPs), which represent singularities embedded in the system's parameter space where the Hamiltonian becomes mathematically defective. Distinct from conventional Hermitian degeneracies, an EP corresponds to a specific critical point in the parameter space where both eigenvalues and eigenstates coalesce, rendering the eigenbasis incomplete~\cite{Bender1998, Heiss2004, Heiss2012, Miri2019}. Beyond their fundamental theoretical intrigue~\cite{Foa_Torres_2018, Longhi_2019F,Tang_2022, Foa_Torres_2020, PYChang2020, CKChiu2021_PRL, Tu2022, Fossati2023, Henry2023, PYYang2024, Zou2024, Ju2025_AP, WeibinLi_2025, Longhih_2025JAP}, EPs have garnered significant attention for their functional utility in quantum technologies. The singular coalescence of eigenstates at the EP leads to a divergent response to external perturbations, a property that has been harnessed to significantly enhance the sensitivity of sensors~\cite{Wiersig2014, Wayne_2017, Hodaei2017, Chen2017, magnondevice_Sci2019, Chen2026nonlinear}, despite ongoing debate\cite{Langbein2018, Lau2018, Zhang2019, Chen2019, Zhong2019, Wang2020, Duggan2022, Ding2023, Naikoo2023, Loughlin2024}. This extreme susceptibility renders non-Hermitian systems a promising platform for next-generation sensing devices, surpassing the fundamental limits of traditional Hermitian schemes.

While conventional EPs in parity-time symmetric systems [i.e., systems with parity-inversion (P) and time-reversal (T) symmetry] are typically associated with fractional power singularities and mark the boundary of PT-symmetry breaking, a novel class of singularities, termed Dirac EPs, has recently been identified~\cite{Rivero2022,Rivero2023,Wu2025_PRL}. Unlike conventional EPs, the Dirac EP can exist entirely within the PT-unbroken phase, maintaining a purely real energy spectrum despite the non-Hermitian coalescence of eigenstates. This singularity draws a compelling analogy to the Dirac cones found in relativistic quantum mechanics and topological phases in Hermitian condensed matter physics, ranging from two-dimensional graphene~\cite{CastroNeto2009_RMP, Novoselov2005, YXJu2024_PRL} to three-dimensional Dirac and Weyl semimetals~\cite{Wang2012, Liu2014, Neupane2014, Armitage2018_RMP, Wan2011, Xu2015, Meng2019, Tzeng2020, Banerjee_2021}. However, a fundamental difference lies in their realization: while these Hermitian Dirac cones reside in the physical momentum space governed by crystal symmetries, the non-Hermitian Dirac EP emerges in a synthetic parameter space. Here, tunable experimental parameters act as effective momenta, generating a linear dispersion relation and unique topological signatures independent of spatial translation symmetry. Following its initial proposal and demonstration in photonic systems~\cite{Rivero2022}, the Dirac EP was recently realized in a solid-state quantum system using a nitrogen-vacancy (NV) center in diamond~\cite{Wu2025_PRL}, opening new avenues for exploring non-Hermitian topology in quantum sensing.

\begin{figure*}[t]
\centering
\includegraphics[width=0.95\textwidth]{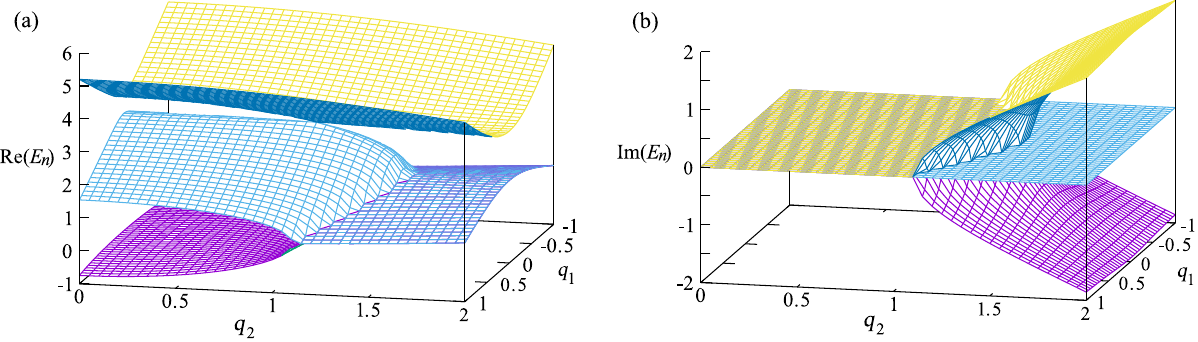}
\caption{\label{fig:energy}The energy spectrum of the non-Hermitian Hamiltonian, Eq.~\eqref{eq:Hamiltonian}. (a) The real part of the eigenenergies, $\mathrm{Re}(E_n)$, revealing a Dirac cone structure centered at the Dirac EP $(q_1,q_2)=(0,1)$ within the PT-unbroken phase. (b) The imaginary part of the eigenenergies, $\mathrm{Im}(E_n)$. The system remains in the PT-unbroken phase characterized by purely real eigenvalues for small $q_2$, but transitions to the PT-broken phase exhibiting non-zero imaginary components when the non-reciprocal coupling exceeds a critical threshold bounded by conventional EPs.}
\end{figure*}

A comprehensive understanding of Dirac EPs demands that we look beyond the spectral degeneracy and examine the structure of the underlying parameter space. This geometry is naturally described by the quantum geometric tensor~\cite{DJZhang2019a, Solnyshkov2021, Ren2024_PRA, Tzeng2021, Tu2023, Hu:24, ChenHe2024}, whose symmetric (metric) and antisymmetric (curvature) sectors encode the quantum metric and Berry curvature of the eigenstate manifold. In the context of quantum information, this metric is intimately related to the fidelity susceptibility~\cite{Tzeng2021, Tu2023}, which characterizes the sensitivity of eigenstates to infinitesimal parameter perturbations. Previous theoretical frameworks have established geometric laws for conventional EPs associated with PT-symmetry breaking. Specifically, analytical lemmas have demonstrated that for a pair of states straddling a phase transition, where one state resides in the PT-unbroken phase and the other in the PT-broken phase, the real part of the fidelity approaches a fixed limit of $1/2$~\cite{Tu2023}. Furthermore, when approaching a conventional EP specifically from the PT-broken phase, the real part of the fidelity susceptibility has been shown to diverge to negative infinity~\cite{Tzeng2021, Tu2023}. However, Dirac EPs present a fundamental challenge to these established results because they reside entirely within the PT-unbroken phase and do not demarcate a transition between broken and unbroken regimes. Consequently, the prerequisite conditions for the previous lemmas, namely either straddling the phase boundary or approaching from the broken phase, are absent. It is thus entirely unknown whether the fidelity approaching a Dirac EP retains any universal feature, or if the unique linear energy dispersion induces a novel geometric anisotropy that defies the behavior of conventional EPs.

In this work, we theoretically investigate fidelity-based geometric signatures of Dirac EPs within a physically realizable solid-state NV-center platform. We first identify these signatures numerically through the fidelity susceptibility across the synthetic parameter space and, subsequently provide a physical interpretation rooted in the defective eigenstate structure of the Dirac EP. Our analysis focuses on an effective non-Hermitian Hamiltonian describing NV centers in diamond, a solid-state system in which Dirac EPs have been experimentally observed~\cite{Wu2025_PRL}. This platform, therefore, provides an ideal and experimentally relevant testbed for exploring the quantum geometry of Dirac EPs.

While theoretical frameworks such as the generalized vielbein formalism have been proposed to map non-Hermitian Hamiltonians onto Hermitian ones via metric deformations~\cite{Ju2022}, the experimental realization considered here relies on the quantum dilation method~\cite{Wu2019_Science, Zhang2021_PRL}. By coupling the spin-1 electronic qutrit to an ancillary nuclear spin, the composite system evolves under a dilated $6\times6$ Hermitian Hamiltonian. Conditional measurements and post-selection on the ancilla state then project an effective $3\times3$ non-Hermitian Hamiltonian, enabling precise engineering of the structure required to host Dirac EPs.

Leveraging this controllable platform, we calculate the fidelity susceptibility across the synthetic parameter space to characterize the geometric properties of the Dirac EP. We demonstrate that the Dirac EP indeed induces a geometric singularity, confirming the universality of fidelity as a probe even entirely within the PT-unbroken phase. Crucially, however, we reveal that this divergence exhibits a singular anisotropy: the susceptibility diverges to negative infinity along the coupling parameter direction while remaining finite along the detuning axis. This behavior stands in stark contrast to conventional EPs, where the geometric singularity is robust against the direction of approach. Our results thus provide a more comprehensive picture of fidelity-based detection, establishing its validity and unique signatures across the full spectrum of non-Hermitian singularities.

\section{Model and Method}
We consider a single nitrogen-vacancy center in diamond, which constitutes a spin-1 qutrit system. To explore the fidelity properties of Dirac EPs, we employ the effective non-Hermitian Hamiltonian experimentally realized via the quantum dilation method~\cite{Wu2025_PRL}. Originating from a truncated tight-binding model with non-reciprocal couplings, the effective Hamiltonian governing the system dynamics is given by
\begin{equation}\label{eq:Hamiltonian}
 H(q_1,q_2)=3S_z^2+2q_1S_z+\sqrt{2}(S_x-iq_2S_y),
\end{equation}
where $S_{x,y,z}$ are the standard spin-1 operators. In this synthetic parameter space, the dimensionless parameters $q_1$ and $q_2$ correspond to the effective momentum and the degree of non-reciprocal coupling, respectively. The interplay between the diagonal terms and the non-Hermitian coupling term $\sqrt{2}(S_x-iq_2S_y)$ generates the specific topological structure required to host the Dirac EP.

Given the non-Hermitian nature of the Hamiltonian where $H \neq H^\dagger$, the left $|L_n\rangle$ and right $|R_n\rangle$ eigenstates are distinct. These states are defined by the eigenvalue equations $\langle L_n|H=E_n\langle L_n|$ and $H|R_n\rangle=E_n|R_n\rangle$, where the band index $n$ takes values of $0$ and $\pm 1$, with $n=0$ denoting the specific state under investigation. Away from the EPs, these eigenstates form a complete biorthogonal basis satisfying the normalization condition $\langle L_n | R_m \rangle = \delta_{nm}$ and the completeness relation
\begin{equation}\label{eq:completeness}
 \sum_n |R_n\rangle \langle L_n| = \openone,
\end{equation}
where $\openone$ is an identity matrix.

The resulting energy spectrum $E_n(q_1, q_2)$ is illustrated in Fig.~\ref{fig:energy}. A key feature of the Dirac EP is its location within the PT-unbroken phase. As shown in Fig.~\ref{fig:energy}(a), the eigenvalues remain purely real in the vicinity of the Dirac EP at $(q_1,q_2)=(0,1)$, where the bands exhibit a linear crossing characteristic of a Dirac cone. This stands in contrast to conventional EPs, which typically mark the boundary of PT-symmetry breaking. However, as the non-reciprocal coupling $q_2$ increases beyond a critical threshold, the system eventually undergoes a transition to the PT-broken phase. As observed in Fig.~\ref{fig:energy}(b), this transition is accompanied by the emergence of conventional EPs (square-root singularities) and complex conjugate eigenenergies. In this work, we focus primarily on the quantum geometry near the Dirac EP in the real-spectrum regime.

It is important to note that the completeness relation Eq.~\eqref{eq:completeness} holds strictly only when the Hamiltonian is diagonalizable. At the exact EP, the eigenstates coalesce, rendering the eigenbasis defective and the resolution of identity invalid. In such defective regimes, a more rigorous treatment involves introducing a metric operator to characterize the curved Hilbert space geometry~\cite{Ju2019, DJZhang2019b, Meden2023_ROPP, Ju2024, Gardas2016, Ju2025_PRA, WMChen2025}. However, the present study focuses on the quantum geometry within the PT-unbroken phase as the system approaches the EP. Since the system parameters remain in the non-defective regime where the spectrum is real and non-degenerate, the biorthogonal formalism remains rigorously well-defined and is consistent with the post-selection measurement scheme employed in experiments.

\begin{figure}[t]
\centering
\includegraphics[width=0.43\textwidth]{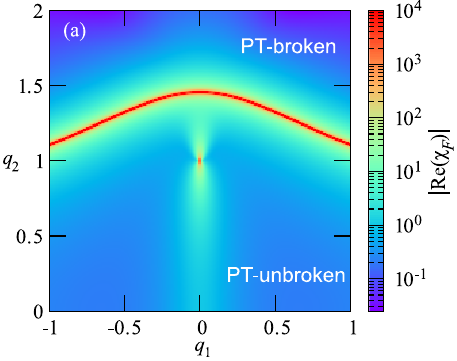}
\includegraphics[width=0.43\textwidth]{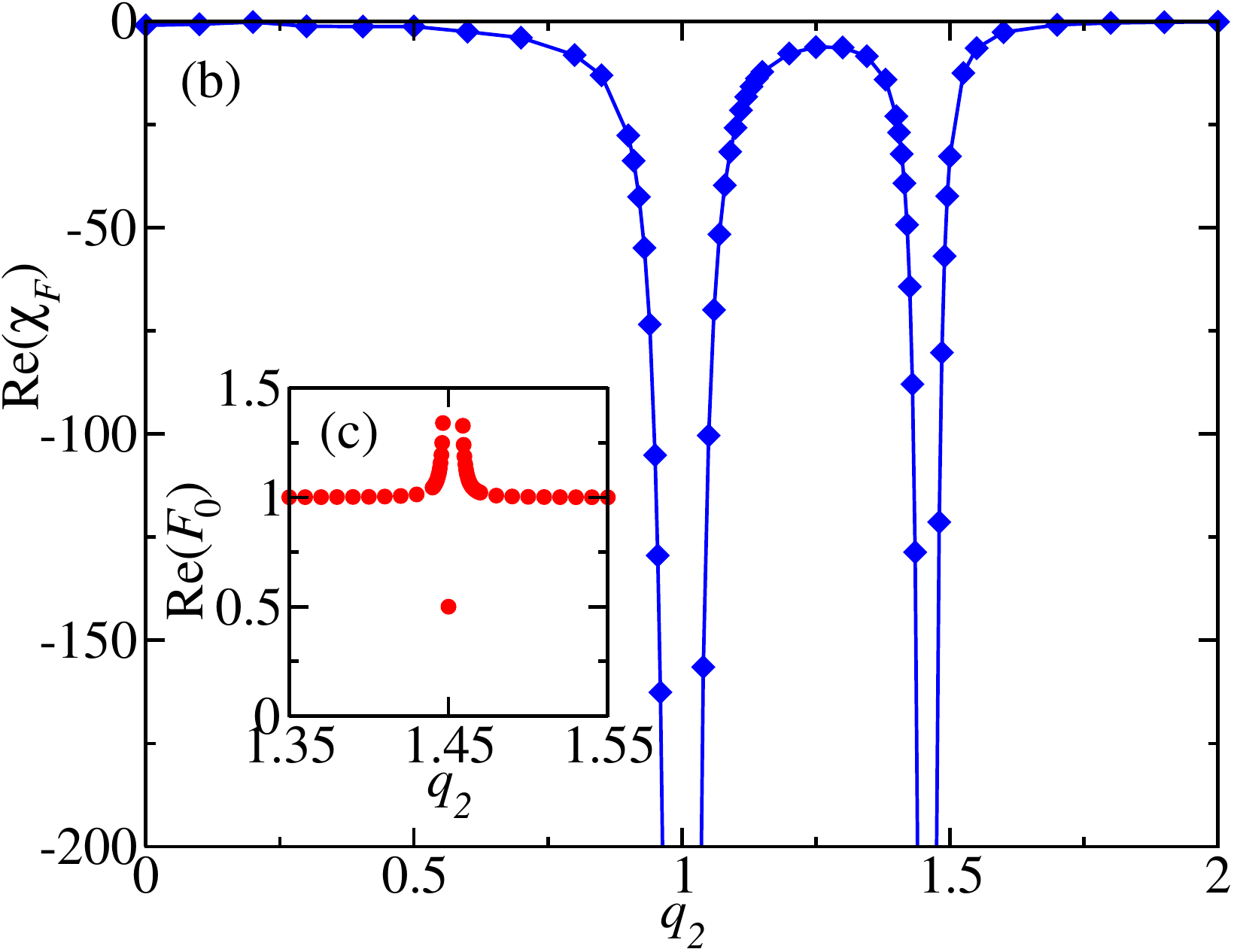}
\caption{\label{fig:Chi0DensityPlot} (a) Density plot of the real part of the fidelity susceptibility $|\mathrm{Re}(\chi_F)|$ in the $(q_1, q_2)$ plane. The red spot at $(0,1)$ corresponds to the Dirac EP, while the red curves indicate the exceptional lines formed by conventional EPs. The displacement direction $\hat n=(0,1)$ is chosen. (b) Behavior of $\mathrm{Re}(\chi_F)$ along the $q_2$ axis with $q_1=0$ fixed, demonstrating sharp divergences to negative infinity at both the Dirac EP and conventional EPs. (c) The real part of the fidelity $F_0$ between states at $q_2$ and $q_2+\delta q$ with fixed $q_1=0$ and $\delta q=10^{-2}$. As the system straddles the conventional EP boundary, $\mathrm{Re}(F_0)$ approaches the universal limit of $1/2$.}
\end{figure}

To elucidate the intrinsic geometry of the Dirac EP, we employ the fidelity susceptibility as a probe in the parameter space. While the definition of fidelity is not strictly unique even within Hermitian quantum mechanics~\cite{Jozsa1994, You2007, MFYang2007, Tzeng2008a, Tzeng2008b, fidelity_review, XJYu2022_PRB}, distinct formulations in the Hermitian context typically yield qualitatively consistent physical predictions regarding phase transitions. For non-Hermitian systems, however, the choice of definition warrants careful consideration because different generalizations can emphasize distinct facets of the underlying quantum geometry~\cite{DJZhang2019a, Solnyshkov2021, Ren2024_PRA, Tzeng2021, Tu2023, Jiang2018, Matsumoto2020, Nishiyama2020, Sun2022, You2025_PRB}. In this study, we adopt the fidelity defined as the complex-valued product of biorthogonal overlaps~\cite{Tzeng2021, Tu2023}. This specific formulation is advantageous as it preserves the metric structure of the Hilbert space and exhibits robust universal properties~\cite{Tu2023, Ju2024, Ju2025_NJP}. Most notably, the real part of the fidelity susceptibility in this formalism diverges to negative infinity, which provides a sharp contrast to the positive infinity divergence characteristic of conventional Hermitian quantum phase transitions. 

Specifically, for a multi-parameter Hamiltonian $H(\vec{q})$ with $\vec{q} = (q_1, q_2)$, the fidelity between the states at $\vec{q}$ and $\vec{q} + \delta \vec{q} = \vec{q} + \delta q \hat{n}$ is defined and expanded as
\begin{align}
 F_n(\vec{q},\vec{q}{+}\delta\vec{q})&=\langle L_n(\vec{q}{+}\delta\vec{q})|R_n(\vec{q})\rangle\langle L_n(\vec{q})|R_n(\vec{q}{+}\delta \vec{q})\rangle\\
 &= 1 - \chi_F^{(n)} \delta q^2 + \mathcal{O}(\delta q^3),
 \label{eq:metricisedFidelity}
\end{align}
where $\hat{n}$ is a unit vector in the parameter space specifying the direction of displacement. The second-order coefficient $\chi_n$ represents the leading non-trivial contribution to the fidelity expansion, defined as
\begin{equation}
 \chi_F^{(n)} (\vec{q}, \hat{n}) = \lim_{\delta q \rightarrow 0} \frac{1 - F_n(\vec{q}, \vec{q} + \delta q \hat{n})}{\delta q^2}.
\end{equation}
This formulation highlights that the fidelity susceptibility is inherently direction-dependent, allowing for a precise characterization of the directional sensitivity to parameter perturbations. Notably, within the PT-unbroken phase, the fidelity remains purely real~\cite{Tu2023}, ensuring that $\chi_F^{(n)}$ remains a real-valued measure throughout the regime where the energy spectrum is real. Hereafter we refer to $\chi_F^{(0)}$ as $\chi_F$.

\section{Numerical Results}\label{sec:results}
Figure~\ref{fig:Chi0DensityPlot}(a) provides a global view of the singular features in the parameter space by mapping the real part of the fidelity susceptibility for the $n=0$ state. Two distinct types of singularities are clearly visible: a localized singularity at $(q_1, q_2)=(0,1)$ identifying the Dirac EP and a continuous boundary of singularities corresponding to the exceptional line composed of conventional EPs. This landscape demonstrates that the fidelity susceptibility serves as a sensitive geometric probe, effectively capturing the distinct signatures of both the novel Dirac EP and the conventional EPs.

To quantitatively examine the divergent behavior, we investigate the parameter space along the $q_2$ axis by setting $q_1 = 0$ and choosing the displacement direction $\hat{n} = (0,1)$, as depicted in Fig.~\ref{fig:Chi0DensityPlot}(b). Along this trajectory, $\mathrm{Re}(\chi_F)$ exhibits sharp singularities, diverging to negative infinity as the system approaches both the Dirac EP and the conventional EPs. This observation confirms that the geometric signatures previously established for non-Hermitian degeneracies~\cite{Tzeng2021,Tu2023} remain valid for Dirac EPs, indicating that the negative divergence of the fidelity susceptibility constitutes a robust and universal hallmark of non-Hermitian singularities.

We further examine the behavior near conventional EPs by considering parameter steps that straddle the exceptional boundary, $q_2 < q_{\mathrm{EP}} < q_2 + \delta q$. As shown in Fig.~\ref{fig:Chi0DensityPlot}(c), the real part of the fidelity $\mathrm{Re}(F_0)$ converges to the universal value of $1/2$ predicted for transitions between PT-unbroken and PT-broken phases~\cite{Tu2023}. Together, these results validate the fidelity susceptibility as a unified geometric probe capable of detecting both conventional and Dirac EPs.

\begin{figure}[t]
\includegraphics[width=0.45\textwidth]{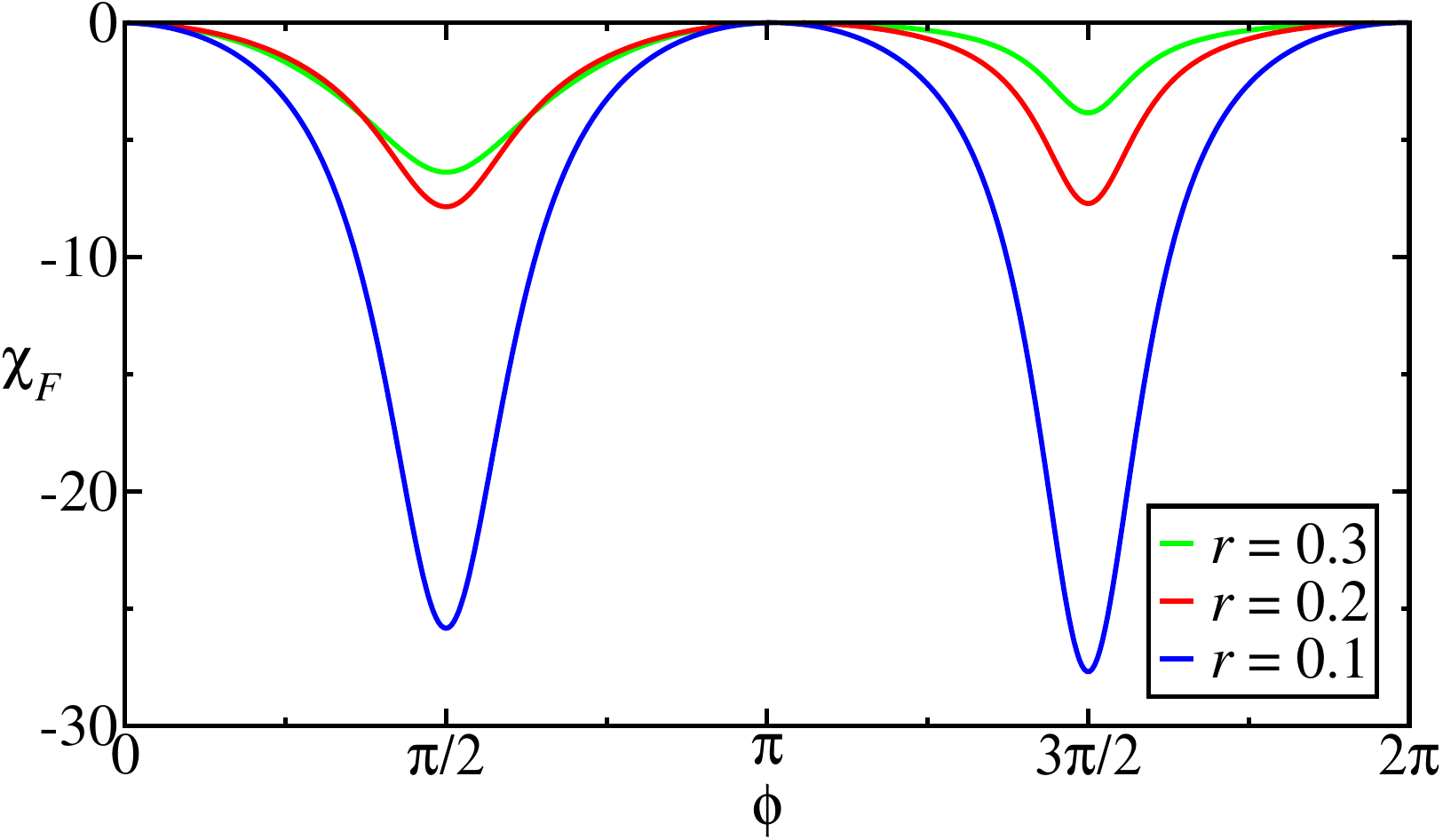}
\caption{\label{fig:Chi0Anisotropy} The real part of the fidelity susceptibility, $\mathrm{Re}(\chi_F)$, calculated along the radial direction as a function of the polar angle $\phi$ surrounding the Dirac EP. The curves correspond to different radial distances $r=0.1, 0.2, 0.3$. The susceptibility exhibits a strong anisotropic divergence, peaking at $\phi=\pi/2$ and $3\pi/2$ (approaching along the $q_2$ axis) while vanishing within numerical precision at $\phi=0$ and $\pi$ (approaching along the $q_1$ axis). This behavior indicates that the leading-order geometric response near the Dirac EP is highly directional in parameter space.}
\end{figure}

Having established this universal behavior, we now turn to a key distinction that sets the Dirac EP apart from conventional EPs: its pronounced directional dependence. To further elucidate the intrinsic geometry of the Dirac EP, we investigate the angular dependence of the fidelity susceptibility in the vicinity of the singularity, as shown in Fig.~\ref{fig:Chi0Anisotropy}. We introduce a polar coordinate system centered at the Dirac EP, defined by $q_1 = r \cos\phi$ and $q_2 = 1 + r \sin\phi$, where $r$ denotes the radial distance and $\phi$ the polar angle. We then compute $\chi_F$ along the radial direction by evaluating the fidelity between states at $r$ and $r - \delta r$ for different values of $\phi$.

Figure~\ref{fig:Chi0Anisotropy} further reveals that the anisotropy of the fidelity susceptibility is not merely quantitative but qualitative: while $\mathrm{Re}(\chi_F)$ diverges strongly for approaches aligned with the $q_2$ axis, it vanishes within numerical precision for approaches along the $q_1$ axis ($\phi = 0,\pi$). As we show below, for the specific Hamiltonian considered here, the symmetry of the perturbation enforces that the leading-order eigenstate deformation along the generalized Jordan direction vanishes for parameter variations aligned with the $q_1$ axis. As a result, the radial fidelity susceptibility is numerically indistinguishable from zero along these directions, whereas it diverges along directions that couple strongly to the defective channel.

\section{Analytic Interpretation: Eigenstate Geometry Near the Dirac Exceptional Point}
\label{sec:analytics}
To elucidate the physical origin of the anisotropic fidelity susceptibility observed in Figs.~\ref{fig:Chi0DensityPlot} and~\ref{fig:Chi0Anisotropy}, we now develop an analytic description of the eigenstate structure in the vicinity of the Dirac EP. While the energy spectrum near the Dirac EP exhibits a linear dispersion reminiscent of Hermitian Dirac cones, the singular geometric response uncovered in Sec.~\ref{sec:results} originates from the defective nature of the underlying eigenstates rather than from spectral gap closing. The following analysis demonstrates how this defectiveness constrains the quantum geometry and leads to a highly directional critical response.

\subsection{Jordan structure at the Dirac exceptional point}
At the Dirac EP located at $(q_1,q_2)=(0,1)$, two eigenvalues of the non-Hermitian Hamiltonian coalesce at energy $E_0=3$, and the Hamiltonian becomes non-diagonalizable. As a result, the EP is characterized by a rank-2 Jordan block rather than a complete set of eigenvectors. One may, therefore, introduce a right Jordan chain $\{|\psi_0\rangle, |\chi\rangle\}$ defined by
\begin{equation}
(H - 3 \openone)|\psi_0\rangle = 0, \qquad
(H - 3 \openone)|\chi\rangle = |\psi_0\rangle,
\end{equation}
together with the corresponding left Jordan chain $\{\langle\phi_0|, \langle\eta|\}$ satisfying
\begin{equation}
\langle\phi_0|(H - 3 \openone) = 0, \qquad
\langle\eta|(H - 3 \openone) = \langle\phi_0|.
\end{equation}
These vectors span the defective subspace associated with the Dirac EP. Away from the EP, where the Hamiltonian remains diagonalizable and the spectrum is real and non-degenerate, the Jordan vectors continuously evolve into the two EP-related eigenstates.

\subsection{Eigenstate expansion near the Dirac EP}
We now consider a small displacement away from the Dirac EP in parameter space,
\begin{equation}
q_1 = r \cos\phi, \qquad q_2 = 1 + r \sin\phi,
\end{equation}
with $r \ll 1$. In contrast to conventional EPs, where eigenstates generally exhibit fractional-power (Puiseux) expansions in the vicinity of the degeneracy~\cite{Heiss2012,Kato1976}, the Dirac EP admits a regular Taylor expansion along real parameter directions.

A key feature of the Dirac EP is the asymmetric response of the two coalescing eigenstates. One branch remains geometrically inert to linear order,
\begin{equation}
|\Psi_-(r,\phi)\rangle = |\psi_0\rangle + \mathcal{O}(r^2),
\end{equation}
while the other branch acquires a linear admixture of the generalized Jordan vector,
\begin{equation}
|\Psi_+(r,\phi)\rangle
=
|\psi_0\rangle + r\,A(\phi)\,|\chi\rangle + \mathcal{O}(r^2).
\end{equation}
Here, the scalar coefficient
\begin{equation}
A(\phi) = \langle \phi_0 | \delta H(\phi) | \psi_0 \rangle
\end{equation}
encodes the projection of the perturbation
$\delta H = r(\cos\phi\,\partial_{q_1}H + \sin\phi\,\partial_{q_2}H)$
onto the defective direction in Hilbert space.

Importantly, this expansion reveals that all leading-order eigenstate deformation near the Dirac EP is confined to a single direction in Hilbert space, namely the generalized Jordan vector $|\chi\rangle$. The EP eigenvector $|\psi_0\rangle$ itself remains unchanged to linear order.

\subsection{Origin of anisotropic fidelity susceptibility}
This constrained eigenstate structure provides a direct explanation for the directional dependence of the fidelity susceptibility observed numerically. Since the biorthogonal fidelity probes the overlap between eigenstates at nearby points in parameter space, its leading nontrivial contribution depends on how strongly the eigenstate changes along the direction of displacement.

Using the expansion above, the fidelity susceptibility of the sensitive branch takes the schematic form
\begin{equation}
\chi_F(\vec{q},\hat{n}) \sim -\frac{|\partial_{\hat n} A(\phi)|^2}{[E_+(r,\phi)-E_-(r,\phi)]^2} + \chi_F^{(\mathrm{reg})},
\end{equation}
where $\hat{n}$ denotes the direction of parameter variation and $\chi_F^{(\mathrm{reg})}$ represents finite contributions arising from non-degenerate bands.

Because the eigenvalue splitting near the Dirac EP is linear and remains nonzero in all real directions, the anisotropic divergence of the fidelity susceptibility is not driven by a vanishing energy gap. Instead, it is governed by the angular dependence of the coefficient $A(\phi)$, which determines how strongly a given parameter displacement couples to the defective eigenstate direction and thus determines the prefactor of the observed $r^{-2}$ divergence. As a result, parameter-space directions that induce a strong projection onto $|\chi\rangle$ give rise to a divergent fidelity susceptibility, while directions orthogonal to this defective channel lead to a strong suppression of the radial fidelity response.

The eigenstate expansion establishes that $A(\phi=0)=0$ at the Dirac EP, so that the leading-order divergence of $\chi_F$ is absent along the $q_1$ direction. This accounts for the pronounced anisotropy visible in Fig.~\ref{fig:Chi0Anisotropy}. However, this expansion is local and does not by itself guarantee that $\chi_F$ vanishes exactly at finite distances from the EP. The exact vanishing that we observe numerically along the full line $q_2=1$ thus warrants further explanation.

The parameter $q_1$ enters the Hamiltonian exclusively through the diagonal term $2q_1 S_z$ so that $\partial_{q_1}H$ is diagonal in the present model. As demonstrated in the Appendix, this diagonal structure implies, through a formal argument based on evolution generators, that the fidelity susceptibility vanishes exactly along at least one full line crossing the EP---consistent with what we observe numerically along $q_2=1$.

Furthermore, this anisotropic behavior is generically expected for Dirac EPs: for any Dirac EP with a sufficient number of accessible parameters, there exists at least one direction in parameter space through the EP along which the fidelity susceptibility vanishes exactly, while perturbations along all other directions lead to divergent behavior. The number of such dark lines may depend on the specific model; the Appendix guarantees at least one.

\subsection{Geometric interpretation and contrast with conventional exceptional points}
The analysis above highlights a fundamental geometric distinction between Dirac EPs and conventional EPs. For conventional EPs associated with PT-symmetry breaking, the geometric singularity is effectively isotropic and closely tied to the square-root closing of the energy gap. In contrast, the Dirac EP resides entirely within the PT-unbroken phase and exhibits a form of geometric criticality without spectral instability, but displaying a strong directional dependence.

The anisotropic divergence of the fidelity susceptibility reflects the fact that the quantum geometry near the Dirac EP is effectively rank-one at leading order: a single generalized Jordan direction in Hilbert space governs the singular response. Parameter-space directions orthogonal to this channel do not induce a first-order eigenstate deformation, even though the underlying geometric structure remains finite and well defined.

This directional sensitivity constitutes a defining geometric signature of the Dirac EP. To place it in a broader context, we now embed the fidelity susceptibility within the framework of the quantum geometric tensor, which provides a unified description of the full geometry of the eigenstate manifold in parameter space.

In the differential limit, the fidelity between nearby eigenstates defines a distance measure on the manifold of biorthogonal quantum states in parameter space, and the fidelity susceptibility corresponds to a directional projection of the associated quadratic form. This quadratic form is governed by the quantum geometric tensor (QGT), which encodes both the metric structure of the eigenstate manifold and its Berry curvature. In non-Hermitian systems the QGT must be constructed from the biorthogonal set of right and left eigenvectors. Following the non-Hermitian generalization introduced in Ref.~\onlinecite{Hu:24}, the tensor for the $n$th band in parameter space $\{q_i\}$ is
\begin{align}
T^{(n)}_{ij} =
 \langle\partial_iL_n|\partial_jR_n\rangle-\langle\partial_iL_n|R_n\rangle\langle L_n|\partial_jR_n\rangle,
\end{align}
where $|R_n\rangle$ and $|L_n\rangle$ denote the right and left eigenvectors satisfying the biorthogonal completeness relation, Eq.\eqref{eq:completeness}, and $\partial_i \equiv \partial/\partial q_i$.

In general, non-Hermitian systems, extracting a quantum metric from $T^{(n)}_{ij}$ is not straightforward: the symmetric part, the real part, and the real-and-symmetric part each provide a distinct and inequivalent definition~\cite{Hu:24,ChenHe2024}, and different fidelity conventions lead to quantum metrics that may differ beyond simple prefactors~\cite{DJZhang2019a, Sun2022, Jiang2018, Tu2023}. In the present case, however, PT symmetry resolves this ambiguity. Within the PT-unbroken phase the eigenvectors are real, so $T^{(n)}_{ij}$ is itself real and symmetric, and all three definitions coincide. The quantum metric is, therefore, unambiguously
defined and our results are independent of this choice.

Expanding the fidelity equation \eqref{eq:metricisedFidelity}, as introduced in Ref.~\onlinecite{Tu2023}, to second order then gives
\begin{equation}
\chi_F^{(n)}(\hat n) = \sum_{i,j}g^{(n)}_{ij}\,\hat n_i \hat n_j ,
\end{equation}
so the fidelity susceptibility directly probes the local metric geometry of the eigenstate manifold. The anisotropic divergence reported in our calculations, therefore, reflects a singular and strongly anisotropic metric structure developing near the Dirac EP.

The antisymmetric part of $T^{(n)}_{ij}$, which in the Hermitian case encodes the Berry curvature, likewise vanishes throughout the PT-unbroken phase, again as a direct consequence of the eigenvectors being real. The anisotropic divergence of the fidelity susceptibility, therefore, has no counterpart in the Berry curvature, and the singular geometry near the Dirac EP is encoded entirely in the quantum metric. We note that in the PT-broken phase the Berry curvature becomes nonzero and exhibits additional asymmetries.
\footnote{A full analysis of the geometric tensor in the vicinity of the Dirac EP and across the phase boundary into the PT broken phase will be presented elsewhere.}

\section{Conclusions}
We have investigated the intrinsic quantum geometry of the Dirac exceptional point within a solid-state NV-center system. By employing the fidelity susceptibility defined as the complex-valued product of biorthogonal overlaps~\cite{Tzeng2021}, which notably remains purely real within the PT-unbroken phase~\cite{Tu2023}, we established a unified geometric characterization of this novel class of non-Hermitian singularities.

A central finding of this study is that the Dirac EP shares a universal geometric hallmark with conventional EPs, namely, the divergence of the real part of the fidelity susceptibility to negative infinity~\cite{Tzeng2021}. This result demonstrates that geometric criticality is a robust feature of non-Hermitian degeneracies and persists even in the absence of a PT-symmetry breaking transition. The consistency of our approach is further corroborated by the analysis of conventional EPs within the same system, where the real part of the fidelity itself converges to the universal limit of $1/2$ when the parameter displacement straddles the exceptional line~\cite{Tu2023}.

Beyond this criticality, we uncovered a pronounced and distinctive geometric anisotropy that can arise naturally near Dirac EPs, and fundamentally differentiates Dirac EPs from their conventional counterparts. While conventional EPs typically exhibit a divergence of the fidelity susceptibility that is insensitive to the direction of approach, the Dirac EP displays a highly directional response. In particular, the fidelity susceptibility diverges to negative infinity along the non-reciprocal coupling axis $q_2$, while being strongly suppressed, and in the present model vanishing identically, along the effective momentum direction $q_1$. Importantly, this anisotropy does not originate from vanishing spectral gaps, as the energy dispersion near the Dirac EP remains linear and nonzero in all real directions. Instead, it reflects the underlying defective eigenstate structure of the Dirac EP, in which the leading-order deformation of the eigenstate is confined to a single generalized Jordan direction in Hilbert space. Strikingly, in the present model, the vanishing of $\chi_F$ along the dark direction is not merely a leading-order suppression but remains exact along the corresponding line through the Dirac EP, as established through a formal argument based on evolution generators (the Appendix). 
More generally, the Appendix proves that at least one such dark line through the Dirac EP is guaranteed to exist for any Dirac EP whose parameter derivatives $\{\partial_{q_i}H\}$ collectively span the off-diagonal operator space at the Dirac EP: since dark directions always come in opposing pairs, the dark locus contains at least one full line through the Dirac EP. The precise number and orientation of dark lines may remain model dependent, but their existence is a provable generic consequence of the Dirac EP structure whenever this spanning condition is satisfied.

These results establish a comprehensive geometric understanding of Dirac EPs by shifting the analytical focus from spectral degeneracies to the structure of eigenstates in parameter space. As a fundamental probe of quantum geometry, the fidelity susceptibility provides a sensitive and experimentally relevant benchmark for quantifying eigenstate response near non-Hermitian singularities. The strong directional dependence revealed in this work demonstrates that geometric criticality at Dirac EPs is governed by specific structural constraints rather than being a uniform property. Consequently, our findings offer clear guidelines for the controlled manipulation of Dirac EPs, indicating that parameter variations must be carefully aligned with the directions of maximal geometric sensitivity to achieve predictable and enhanced responses. The Appendix further suggests that Dirac EPs may influence the anisotropy of the quantum metric beyond the immediate vicinity of the exceptional point, motivating future studies of their broader impact on non-Hermitian phase diagrams. More broadly, this work establishes fidelity-based diagnostics as a powerful framework for exploring anisotropic geometric singularities in a wide range of emerging non-Hermitian platforms and engineered quantum materials.

Finally, we note that the fidelity-based geometric diagnostics discussed here are, in principle, accessible in several experimental platforms that realize non-Hermitian Hamiltonians through conditional dynamics. In the NV-center implementation of Ref.~\onlinecite{Wu2025_PRL}, the effective non-Hermitian evolution is engineered through quantum dilation, enabling direct access to biorthogonal state overlaps via conditional measurements and post-selection. More generally, platforms based on post-selected quantum trajectories, such as the setting considered in Ref.~\onlinecite{Rivero2022}, also provide access to pure conditional states, from which state overlaps and fidelity susceptibilities may be inferred through repeated state preparation and measurement. Another potentially suitable platform is provided by trapped-ion quantum simulators~\cite{Ding2021:trapped_ion, Cao2023:trapped_ion, chen2025:trapped_ion}. Although a direct realization of a Dirac EP in a trapped-ion platform remains to be demonstrated, the essential ingredients required for the present proposal are available in principle: a three-level internal manifold can encode the effective qutrit, tunable coherent couplings can reproduce the spin-1 structure of Hamiltonian Eq.~\eqref{eq:Hamiltonian}, and ancilla-assisted dilation or state-selective dissipation can be used to engineer the corresponding non-Hermitian dynamics. Combined with the high-fidelity state tomography routinely available in trapped-ion experiments, these capabilities would provide a natural route to reconstruct the relevant state overlaps and probe the directional fidelity response predicted here. These considerations suggest that fidelity-based probes offer a broadly applicable route to experimentally exploring the intrinsic geometry of non-Hermitian singularities.

\begin{acknowledgments}
YCT thanks Wayne Cheng-Wei Huang and David Mikolas for useful discussions about NV centers.
CYJ and YCT are grateful for the support of the National Science and Technology Council (NSTC) of Taiwan under grant No. NSTC 112-2112-M-110-013-MY3, 114-2112-M-029-003 and 113-2112-M-A49-015-MY3, respectively.
We thank the National Center for High-performance Computing (NCHC) of National Applied Research Laboratories (NARLabs) in Taiwan for providing computational and storage resources. This work is funded in part by a QuantEmX award from the Institute of Complex Adaptive Matter (ICAM) and the Gordon and Betty Moore Foundation through Grant GBMF9616 to Gunnar M\"oller. GM also acknowledges the hospitality of the National Centre for Theoretical Sciences, Taiwan. 
\end{acknowledgments}

\begin{appendix}
 \section{Relation Between Vanishing Fidelity Susceptibilities and Dirac Exceptional Points \label{Appendix:VanishingSusceptibility}}

		In this appendix, we develop a formal argument to understand the anisotropic behavior of the fidelity susceptibility around Dirac exceptional points. Specifically, we explain why the fidelity susceptibility vanishes along a particular direction of approach to the Dirac EP, despite the fact that it is generally expected to diverge~\cite{Tzeng2021, Tu2023, Ju2025_NJP}.

		To proceed, we first state some essential results for parameter-induced evolution generators~\cite{Ju2024}.

		\subsection{Brief Review of Evolution Generators}

			It has been shown that if the Hamiltonian depends on a parameter $q_i$ [i.e., $H = H(q_1, q_2, \dots)$], the evolution of the quantum state $\ket{\psi}$ with respect to $q_i$ is governed by
			\begin{align}
				\partial_i \ket{\Psi} = - i K_i \ket{\Psi},
			\end{align}
			where $\partial_i \equiv \dfrac{\partial}{\partial q_i}$ and $K_i$ is the evolution generator related to the Hamiltonian $H$ by
			\begin{align}
				\frac{\partial}{\partial t} K_i = i \left[K_i, H\right] + \partial_i H. \label{App:EvolutionGeneratorEquation}
			\end{align}

			Additionally, the generators satisfy the following compatibility condition:
			\begin{align}
				\partial_i K_j - \partial_j K_i - i \left[K_j, K_i\right] = 0.
			\end{align}

			The generators are not uniquely determined. The freedom remaining is, in fact, a result of gauge symmetry. To simplify the discussion of fidelity susceptibility, it is standard practice to adopt the adiabatic gauge, which satisfies the condition
			\begin{align}
				\left[\frac{\partial}{\partial t} K_i, H\right] = 0. \label{App:AdiabaticGauge}
			\end{align}

			A direct consequence of the adiabatic gauge is that the generators are at most linear in time, i.e.,
			\begin{align}
				K_i = t \p{K_i}{1} + \p{K_i}{0},
			\end{align}
			where $\p{K_i}{1}$ and $\p{K_i}{0}$ are time-independent operators. Moreover, the differential equations in Eqs.~\eqref{App:EvolutionGeneratorEquation} and \eqref{App:AdiabaticGauge} simplify to algebraic equations,
			\begin{align}
				& \p{K_i}{1} = i \left[\p{K_i}{0}, H\right] + \partial_i H,\label{App:AlgebraicForm}\\
				& \left[\p{K_i}{1}, H\right] = 0. \label{App:Commute}
			\end{align}

			In addition, the evolution of the left- and right-eigenstates in $q_i$ is
			\begin{align}
				& \partial_i \bra{L_n} = i \bra{L_n} \p{K_i}{0},\\
				& \partial_i \ket{R_n} = - i \p{K_i}{0} \ket{R_n}.
			\end{align}

			Interestingly, Eqs.~\eqref{App:AlgebraicForm} and \eqref{App:Commute} imply that $\p{K_i}{1}$ and $H$ share a set of eigenstates. Furthermore, eigenvalues of $\p{K_i}{1}$ are exactly the $q_i$-derivatives of the corresponding Hamiltonian eigenvalues, namely,
			\begin{align}
				& \bra{L_n} \p{K_i}{1} = \left(\partial_i h_n\right) \bra{L_n}, \label{App:LeftEigenstate}\\
				& \p{K_i}{1} \ket{R_n} = \left(\partial_i h_n\right) \ket{R_n}, \label{App:RightEigenstate}
			\end{align}
			where $h_n$ denotes the eigenvalue of $H$ corresponding to the $n$th left/right eigenstate.

			For evolution along a path $\gamma$ parameterized by $s$, where $\gamma(s) = (q_1(s), q_2(s), \dots)$, the evolution equations~\cite{Ju2025} for the left and right eigenstates are given by
			\begin{align}
				& \frac{d}{ds} \bra{L_n} = \sum_i \frac{d q_i}{d s} \partial_i \bra{L_n} = i \sum_i \frac{d q_i}{d s} \bra{L_n} \p{K_i}{0},\\
				& \frac{d}{ds} \ket{R_n} = \sum_i \frac{d q_i}{d s} \partial_i \ket{R_n} = - i \sum_i \frac{d q_i}{d s} \p{K_i}{0} \ket{R_n}.
			\end{align}
			By defining the effective generator along the path as
			\begin{align}
				\p{K}{0} = \sum_i \frac{d q_i}{d s} \p{K_i}{0},
			\end{align}
			these expressions simplify to
			\begin{align}
				& \frac{d}{ds} \bra{L_n} = i \bra{L_n} \p{K}{0},\\
				& \frac{d}{ds} \ket{R_n} = - i \p{K}{0} \ket{R_n}.
			\end{align}

			Furthermore, from the linearity in Eq.~\eqref{App:AlgebraicForm}, \eqref{App:Commute}, \eqref{App:LeftEigenstate}, and \eqref{App:RightEigenstate}, we find that
			\begin{align}
				\p{K}{1} = s \sum_i n_i \p{K_i}{1},
			\end{align}
			which satisfies the following relations:
			\begin{align}
				& \p{K}{1} = i \left[\p{K}{0}, H\right] + \partial_s H, \label{App:MasterEq1}\\
				& [\p{K}{1}, H] = 0, \label{App:MasterEq2}\\
				& \bra{L_n} \p{K}{1} = \left(\frac{d}{ds} h_n\right) \bra{L_n}, \label{App:MasterEq3}\\
				& \p{K}{1} \ket{R_n} = \left(\frac{d}{ds} h_n\right) \ket{R_n}. \label{App:MasterEq4}
			\end{align}

			In other words, the system behaves as if the state were evolving with respect to the single parameter $s$.

			The final element required for the current discussion is the fidelity susceptibility for the $n$th eigenstate along the path $s$. Under the adiabatic gauge defined in Eq.~\eqref{App:AdiabaticGauge}, this susceptibility simplifies to
			\begin{align}
				\p{\chi_F}{n} = \bra{L_n} \left(\p{K}{0}\right)^2 \ket{R_n} - \left(\bra{L_n} \p{K}{0} \ket{R_n}\right)^2. \label{App:FidelitySusceptibility}
			\end{align}

			A comprehensive treatment of the evolution generators, the adiabatic gauge conditions, and the detailed derivation of the fidelity susceptibility presented above can be found in Refs.~\cite{Ju2024, Ju2025}.

		\subsection{Formal Study of Fidelity Susceptibility}

			In this subsection, we consider a Hamiltonian $H = H\left( \vec{q} \right)$ that is linear in the vector $\vec{q} = (q_1, q_2, \dots)$ near an EP at $\vec{q}_\text{\tiny EP} = (0, 0, \dots)$, i.e.,
			\begin{align}
				H & = \p{H}{0} + \sum_i q_i \p{H_i}{1} \label{App:Hamiltonian}\\
				& = \p{H}{0} + q_1 \p{H_1}{1} + q_2 \p{H_2}{1} + \cdots,
			\end{align}
			where $\p{H}{0}$ is at an EP and $\p{H_i}{1}$ are independent of $\vec{q}$.

			We now investigate a perturbation of the left/right eigenvectors $\bra{\p{L_n}{0}}$ and $\ket{\p{R_n}{0}}$ of $\p{H}{0}$ by shifting the parameter vector along a specific direction $\hat{n}$, such that $\vec{q} = s \hat{n} = (s n_1, s n_2, \dots)$ with a small $s$. The corresponding perturbed eigenvalue $h_n$ is expected to follow a Puiseux series expansion of the form
			\begin{align}
				h_n = h_n^{(0)} + s^{p_1} h_n^{(1)} + \cdots. \label{App:EigenvalueExpansion}
			\end{align}
			The exponent $p_1$ in Eq.~\eqref{App:EigenvalueExpansion} is typically a positive fraction in the vicinity of an EP.

			Nevertheless, we now examine the case of vanishing fidelity susceptibility observed in this study. From Eq.~\eqref{App:FidelitySusceptibility}, we have
			\begin{align}
				0 = \p{\chi_F}{n} = \bra{L_n} \left(\p{K}{0}\right)^2 \ket{R_n} - \left(\bra{L_n} \p{K}{0} \ket{R_n}\right)^2, \label{App:VanishingFidelitySusceptibility}
			\end{align}
			where
			\begin{align}
				& \bra{L_n} H = h_n \bra{L_n},\\
				& H \ket{R_n} = h_n \ket{R_n},\\
				& \p{K}{0} = s \sum_i n_i \p{K_i}{0}.
			\end{align}

			An obvious solution to Eq.~\eqref{App:VanishingFidelitySusceptibility} arises if at least one of the states, either $\bra{L_n}$ or $\ket{R_n}$, is also an eigenstate of $\p{K}{0}$. For simplicity, we shall focus on the case where $\ket{R_n}$ is a right eigenstate of $\p{K}{0}$, i.e.,
			\begin{align}
				\p{K}{0} \ket{R_n} = k_n \ket{R_n}.
			\end{align}

			With Eqs.~\eqref{App:MasterEq1} and \eqref{App:MasterEq4}, we find
			\begin{align}
				\left(\frac{d}{ds} h_n\right) \ket{R_n} & = \p{K}{1} \ket{R_n}\\
				& = \left(i \left[\p{K}{0}, H\right] + \partial_s H\right) \ket{R_n}\\
				& = \left(\partial_s H\right) \ket{R_n},
			\end{align}
			where the final equality follows because $\ket{R_n}$ is a common eigenstate of both $\p{K}{0}$ and $H$. Note that this result holds for this specific state, even though $\left[\p{K}{0}, H\right] \neq 0$ in general.

			By substituting the eigenvalue expansion from Eq.~\eqref{App:EigenvalueExpansion} into the equation above, we obtain
			\begin{align}
				\left(\partial_s H\right) \ket{R_n} & = \frac{d}{ds} \left( h_n^{(0)} + s^{p_1} h_n^{(1)} + \cdots \right) \ket{R_n}\\
				& = \left(p_1 s^{p_1 - 1} + \cdots\right) \ket{R_n}.
			\end{align}
			In other words, $\left(p_1 s^{p_1 - 1} + \cdots\right)$ is an eigenvalue of $\partial_s H$.

			However, because $\partial_s H = \left(n_1 \p{H_1}{1} + n_2 \p{H_2}{1} + \cdots \right)$ is independent of $s$, the eigenvalue must also be independent of $s$. This implies that $p_1 = 1$ and that all higher-order terms in the expansion of $h$ vanish, resulting in the linear form
			\begin{align}
				h = \p{h}{0} + s \p{h}{1}.
			\end{align}

			In fact, this conclusion also holds if the alternative condition is satisfied, namely, that $\bra{L_n}$ as a left eigenstate of $\p{K}{0}$. The derivation proceeds analogously to the previous discussion by replacing the right eigenstate $\ket{R_n}$ with the left eigenstate $\bra{L_n}$.

			In other words, the fidelity susceptibility $\p{\chi_F}{n}$ vanishes only when the perturbation is applied along a direction that reduces the eigenvalue expansion to a linear dependence on the perturbation parameter. Given that this linearity is a defining characteristic of Dirac EPs, the primary analytical concern reduces to the existence of a unit vector $\hat{n}$ such that the Hamiltonian eigenstate $\ket{R_n}$ (or $\bra{L_n}$) is also an eigenstate of $\p{K}{0} = \sum_i n_i \p{K_i}{0}$.

				To show that such an $\hat{n}$ exists, we first construct an arbitrary matrix $\Xi$ that renders either $\bra{L_n}$ or $\ket{R_n}$ a left or right eigenstate, respectively. Without loss of generality, we choose to work with $\ket{R_n}$.

			To be more specific, we choose a matrix $\Xi$ such that
			\begin{align}
				\Xi \ket{R_n} = \xi \ket{R_n}, \label{App:ArbitraryMatrix}
			\end{align}
			where $\xi$ represents the corresponding eigenvalue.

				Since $\p{K_i}{0}$ associated with the $q_i$-direction can be derived from Eq.~\eqref{App:AlgebraicForm}, namely,
			\begin{align}
				\p{K_i}{1} = i \left[\p{K_i}{0}, H\right] + \p{H_i}{1}
			\end{align}
			these matrices are linearly independent, provided that the corresponding $\p{H_i}{1}$ (and $\p{H}{0}$) are linearly independent.

			In other words, with a sufficient number of $\p{K_i}{0}$ (i.e., sufficiently many parameters, as further defined below), we can always decompose $\Xi$ into the superposition of $\p{K_i}{0}$ and a matrix $\Lambda$ that commute with $H$, namely,
			\begin{align}
				\Xi = \Lambda + \sum_i c_i \p{K_i}{0}. \label{App:CompletenessMatrix}
			\end{align}
			To show this, we first note that the Hamiltonian here is not at an EP but is ``$\vec{q}$'' away from it [see Eq.~\eqref{App:Hamiltonian}], although it remains very close to the EP. Consequently, the left and right eigenstates form a complete basis. Since both $\p{K_i}{1}$ and $\Lambda$ commute with the Hamiltonian $H$, they can be decomposed as
			\begin{align}
				& \p{K_i}{1} = \sum_{h_p = h_q} \alpha_{ipq} \ket{R_p}\bra{L_q},\\
				& \Lambda = \sum_{h_p = h_q} \beta_{pq} \ket{R_p}\bra{L_q},
			\end{align}
			while $\partial_i H$, $\p{K_i}{0}$, and $\Xi$ take the form
			\begin{align}
				& \p{H_i}{1} = \sum_{p, q} \gamma_{ipq} \ket{R_p}\bra{L_q},\\
				& \p{K_i}{0} = \sum_{p, q} \zeta_{ipq} \ket{R_p}\bra{L_q},\\
				& \Xi = \sum_{p, q} \eta_{ipq} \ket{R_p}\bra{L_q},
			\end{align}
			where $\gamma_{ipq}$ is determined by the Hamiltonian and the parameter $q_i$.

             More specifically, with the $\p{K_i}{0}$ derived from the parameter derivatives 
$\p{H_i}{1}$, the decomposition Eq.~\eqref{App:CompletenessMatrix} is achievable 
provided that the off diagonal matrix elements $\langle L_p | \p{H_i}{1} | R_q \rangle$ 
for $h_p \neq h_q$, taken collectively over all parameters $q_i$, span the full space 
of off diagonal operators in the eigenbasis of $H$. This is the precise condition 
defining the informally expressed requirement of ``sufficiently many parameters.''

			For each $q_i$, Eq.~\eqref{App:AlgebraicForm} becomes
			\begin{align}
				& \sum_{h_p = h_q} \alpha_{ipq} \ket{R_p}\bra{L_q}\\
				& = i \left[\sum_{p, q} \zeta_{ipq} \ket{R_p}\bra{L_q}, H\right] + \sum_{p, q} \gamma_{ipq} \ket{R_p}\bra{L_q}\\
				& = \sum_{h_p \neq h_q} (h_q - h_p) \zeta_{ipq} \ket{R_p}\bra{L_q} + \sum_{p, q} \gamma_{ipq} \ket{R_p}\bra{L_q}.
			\end{align}
			Comparing both sides of the equation, we obtain
			\begin{align}
				\left\{
					\begin{array}{ll}
						\alpha_{ipq} = \gamma_{ipq}, & \text{for $h_p = h_q$}\vspace{0.1cm}\\
						\zeta_{ipq} = \dfrac{\gamma_{ipq}}{h_p - h_q}, & \text{for $h_p \neq h_q$}
					\end{array}
				\right..
			\end{align}
			Therefore, we can rewrite $\p{K_i}{0}$ as
			\begin{align}
				\p{K_i}{0} & = \sum_{h_p = h_q} \zeta_{ipq} \ket{R_p}\bra{L_q} + \sum_{h_p \neq h_q} \zeta_{ipq} \ket{R_p}\bra{L_q}\\
				& = \sum_{h_p = h_q} \zeta_{ipq} \ket{R_p}\bra{L_q} + \sum_{h_p \neq h_q} \frac{\gamma_{ipq}}{h_p - h_q} \ket{R_p}\bra{L_q}.
			\end{align}

			In other words, Eq.~\eqref{App:CompletenessMatrix} becomes
			\begin{align}
				\begin{split}
					\sum_{p, q} \eta_{ipq} \ket{R_p}\bra{L_q} = & \sum_{h_p = h_q} (\beta_{pq} + \sum_i c_i \zeta_{ipq}) \ket{R_p}\bra{L_q}\\
					& + \sum_i c_i \sum_{h_p \neq h_q} \frac{\gamma_{ipq}}{h_p - h_q} \ket{R_p}\bra{L_q},
				\end{split}
			\end{align}
			which leads to
			\begin{align}
				\left\{
					\begin{array}{ll}
						\displaystyle \eta_{pq} = \beta_{pq} + \sum_i c_i \zeta_{ipq}, & \text{for $h_p = h_q$}\vspace{0.1cm}\\
						\displaystyle \eta_{pq} = \sum_i \dfrac{c_i \gamma_{ipq}}{h_p - h_q}, & \text{for $h_p \neq h_q$}
					\end{array}
				\right.. \label{App:Completion}
			\end{align}

			Therefore, if $\p{H_i}{1}$ and $\p{H}{0}$ are linearly independent and we have sufficiently many $\p{H_i}{1}$ (or $\gamma_{ipq}$), we can always find a set of $c_i$ that renders Eq.~\eqref{App:Completion} valid.

			Consequently, Eq.~\eqref{App:ArbitraryMatrix} can be expressed as
			\begin{align}
				\left(\Lambda + \sum_i c_i \p{K_i}{0}\right) \ket{R_n} = \xi \ket{R_n}.
			\end{align}
			Given that $\Lambda$ commutes with $H$, it shares the same eigenstate basis, allowing us to simplify the expression to
			\begin{align}
				\left(\sum_i c_i \p{K_i}{0}\right) \ket{R_n} = \left(\xi - \lambda\right) \ket{R_n},
			\end{align}
			where $\lambda$ denotes the eigenvalue of $\Lambda$ associated with the eigenstate $\ket{R_n}$.

			Since the vector of coefficients $\vec{c} = (c_1, c_2, \dots)$ is defined up to an arbitrary global scaling factor, we can normalize the expression by dividing both sides by its norm, $|\vec{c}| = \sqrt{\sum_i |c_i|^2}$, to obtain
			\begin{align}
				\left(\sum_i \frac{c_i}{|\vec{c}|} \p{K_i}{0}\right) \ket{R_n} = \frac{\xi - \lambda}{|\vec{c}|} \ket{R_n}.
			\end{align}
			This rescaling does not alter the fact that $\ket{R_n}$ remains an eigenstate of the operator. Consequently, by defining the unit vector $\hat{n} = \frac{\vec{c}}{|\vec{c}|}$, this construction shows that such a direction generically exists. Since the construction begins with an arbitrary choice of $\Xi$, replacing $\Xi \rightarrow - \Xi$ is equally valid and yields $- \hat{n}$ as a dark direction whenever $\hat{n}$ is. Dark directions, therefore, always come in opposing pairs, guaranteeing that the dark locus contains at least one full line through the Dirac EP. The evolution-generator argument further shows that this vanishing can extend along a line through the Dirac EP and to finite distances from it, provided that the eigenvalue depends linearly on the corresponding perturbation parameter.

			Given that the eigenvalue structure of Dirac EPs is inherently consistent with the requirement $\chi_F = 0$, and that there exists a direction $\hat{n}$ satisfying this condition, we conclude that the vanishing of the fidelity susceptibility is a characteristic property of Dirac EPs. Furthermore, since perturbations along other directions typically lead to a divergence in $\chi_F$, we can identify this pronounced anisotropy as a fundamental feature of Dirac EPs.

\end{appendix}
%

\end{document}